\documentclass[12pt,preprint]{aastex}

\def\ltsima{$\; \buildrel < \over \sim \;$}
\def\gtsima{$\; \buildrel > \over \sim \;$}
\def\lsim{\lower.5ex\hbox{\ltsima}}
\def\gsim{\lower.5ex\hbox{\gtsima}}
\def\lapp{\ifmmode\stackrel{<}{_{\sim}}\else$\stackrel{<}{_{\sim}}$\fi}
\def\gapp{\ifmmode\stackrel{>}{_{\sim}}\else$\stackrel{<}{_{\sim}}$\fi}
\def\mcom{M_{\rm COM}}
\def\mpsr{M_{\rm PSR}}
\def\apsr{a_{\rm PSR}}

\def\Msun{M_{\odot}}
\def\Rsun{R_{\odot}}

\def\psr{PSR\, J1439-5501}
\def\com{ the companion}

\newdimen\minuswidth    

\setbox0=\hbox{$-$}

\minuswidth=\wd0

\catcode`@=\active

\def@{\kern\minuswidth}

\setbox0=\hbox{\rm0}

\shorttitle{Optical companion to PSR J1439-5501}

\shortauthors{Pallanca et al.}

\begin{document} 

\title{The optical companion to the intermediate mass millisecond pulsar J1439-5501 in the
Galactic field\footnote{Based on FORS2 observations collected at the ESO-VLT under program 383.D-0406(A).}}

\author{
C. Pallanca\altaffilmark{2},
B. Lanzoni\altaffilmark{2},
E. Dalessandro\altaffilmark{2},
F. R. Ferraro\altaffilmark{2},
A. Possenti\altaffilmark{3},
M. Salaris\altaffilmark{4},
M. Burgay\altaffilmark{3}.
}
\affil{\altaffilmark{2} Dipartimento di Fisica e Astronomia, Universit\`a degli Studi
di Bologna, Viale Berti Pichat 6/2, I--40127 Bologna, Italy}

\affil{\altaffilmark{3} INAF-Osservatorio Astronomico di Cagliari,
  localit\`a Poggio dei Pini, strada 54, I-09012 Capoterra, Italy}

\affil{\altaffilmark{4} Astrophysics Research Institute, Liverpool John Moores University, UK}

\date{May 16, 2013}
\begin{abstract}

We present the identification of the  companion star to the intermediate mass binary pulsar J1439-5501
obtained by means of ground-based deep images in the $B$, $V$ and $I$ bands, acquired with FORS2 mounted at the ESO-VLT.  
The companion is a massive  white dwarf (WD) with $B=23.57 \pm 0.02 $, $V=23.21 \pm 0.01$ and $I=22.96 \pm 0.01$, located at only $\sim0.05\arcsec$ from the pulsar  radio  position.  
Comparing the WD location in the ($B$, $B-V$) and ($V$, $V-I$) Color-Magnitude diagrams with 
theoretical cooling sequences we derived a range of plausible combinations 
of companion  masses ($1 \lapp \mcom \lapp 1.3~\Msun$), distances ($d \lapp 1200$ pc), radii ($\lapp7.8~10^{-3}~\Rsun$) 
and temperatures ($T=31350^{+21500}_{-7400}$). 
From the PSR mass function and the estimated  mass range we also constrained the inclination angle $i \gapp 55 ^{\circ}$ and the pulsar mass 
($\mpsr \lapp2.2 \Msun$).
The comparison between the WD cooling age and the spin down age suggests that the latter is overestimated by a factor of about  ten.

\end{abstract} 


\section{Introduction}
\label{Sec:intro}
 
According to the canonical recycling scenario, millisecond pulsars (PSRs)
 form in binary systems containing a neutron star (NS) which is
eventually spun up through mass accretion from the evolving
companion. 

Depending on the companion mass it is possible to classify radio PSRs 
with evolved companions into three groups: low-, intermediate- 
and high-mass binary PSRs. In the context of stellar 
evolution this classification is related to the evolution of stars with 
low, intermediate and high mass, that evolve to Helium (He), Carbon-Oxygen (CO)
or Oxigen-Neon-Magnesium (ONeMg) white dwarfs (WDs), and NS, respectively 
(Van Kerkwijk \& Kulkarni, 1995).

In particular intermediate mass binary PSRs (IMBPs) are thought to be 
generated from intermediate mass X-ray Binaries (IMXBs), with donor 
masses typically of $3-6~\Msun$. 
Following a supernova explosion, the primary becomes a NS and 
later the secondary evolves and recycles the PSR through mass transfer, 
eventually forming a CO- or ONeMg-WD with a He envelope
(van den Heuvel 1994; Tauris et al., 2000; Taam et al. 2000).
A massive WD can also be left if the companion to
the PSR evolved up to the asymptotic giant branch before overflowing its 
Roche lobe. The latter scenario may lead to more accretion,
and thus to a NS spun up to faster periods and with 
a more strongly reduced magnetic field. The WD
 might still have a hydrogen envelope. 
Seventeen candidate IMBP systems are currently known in the Galactic field 
(Van Kerkwijk et al., 2005; Jacoby et al 2006; Tauris et al., 2012; Burgay et al., 2012).
 
The optical identification of companion stars to such objects is a crucial point for
unveiling their nature and for investigating the formation channels of
the binary system (Ferraro et al. 2001, 2003, Cocozza et al. 2008, and Pallanca et al.,
2010, 2012). 
In principle, it allows to
accurately estimate the companion mass (see, e.g.,  Antoniadis et al., 2012)
and, in turn, the PSR mass. In that case, binary PSRs can
become a unique laboratory for fundamental physics, where the equation
of state of matter at nuclear densities can be constrained (see e.g. van
Kerkwijk, Breton \& Kulkarni 2011) and gravity theories tested (see, e.g.,
Freire et al., 2012).
In case of the identification of a WD companion, its color and  
magnitude could be used to infer the mass and 
the cooling age of the WD, which could be compared with the age of the 
NS to understand the timescale of the magnetic field decay.
Note however that only 5 companions to IMBPs have been detected so far in the
optical band: PSR B0655+64, PSR J1022+1001, PSR J1528-3146, PSR J1757-5322 and PSR
J2145-0750 (van Kerkwijk et al. 1995, 2005; Lundgren et al. 1996a, 1996b; 
Lohmer et al., 2004  and Jacoby et al. 2006). 

In this work we have collected and analyzed a sample of deep multi band 
images in the direction of \psr\ in the Galactic field, with the aim of identifying the
companion star.  The plan of the paper is as follows: in Section~2 the known properties of
the PSR are listed, while the observations and data analysis are described in Section~3.
Results are presented in Section~4 and discussed in Section~5.

\section{\psr}
\label{Sec:pulsar}

The binary \psr\ was discovered during the first reprocessing (Faulkner et al., 2004) of the 
data of the Parkes Multibeam Pulsar Survey (Manchester et al., 2001) and its accurate timing 
parameters were published by Lorimer et al. (2006, hereafter L06).
The spin period ($P\sim 29$ ms) and the surface magnetic
field ($B_{s}\sim 2.05\times10^9$ G) suggest that \psr\ belongs to
the class of the {\it mildly} recycled PSRs, i.e. the NS that
underwent a relatively short phase of mass accretion from the
companion star.

The distance of \psr\ can be estimated from its dispersion
measure (DM$\sim 15$ pc cm$^{-3}$, L06) once a model for the
distribution of the electrons in the interstellar medium is adopted. In
particular, the TC93 model (Taylor \& Cordes 1993) predicts 760 pc,
the NE2001 model (Cordes \& Lazio 2002) places the PSR at a
 smaller distance (600 pc), whereas applying the updated Taylor \& Cordes model (Schnitzeler, 2012) 
we obtained a larger distance, of 950 pc.   
 Considering  these uncertainties, a possible range\footnote{ However, for our detailed analysis below, the range 
of available WD models forces us to use a slightly reduced range, $d= 600-1200$ pc 
(but see section \ref{Sec:Results}).} for the distance is $d= 500-1200$ pc.

From the observed orbital period, $P_b = 2.117942520(3)$ d, and the
projected semi-major axis of the orbit, $\apsr\sin
i=2.947980(3)\times 10^{11}$ cm $\sim 4.2$ $\Rsun$ (where $i$ is the
orbital inclination), L06 derived a PSR mass
function $f_{\rm PSR}=0.227597~\Msun$. This value
implies that the companion to \psr\ has a mass $\gapp 1~\Msun$. In
particular, adopting a NS mass $\mpsr = 1.4~\Msun$, the
minimum companion mass is $1.13~\Msun$ (L06), while assuming the
minimum value for  radio PSR mass measured so far 
($\mpsr = 1.24~\Msun$, Faulkner et al., 2005), the minimum companion mass
would be $1.07~\Msun.$
 
The minimum value for the companion mass leaves 3 options for the nature of
this object: an ordinary non degenerate star, a NS, or a very
massive CO- or ONe-WD.  
However a Main Sequence (MS) star with mass as large as $1.1~\Msun$ 
located at the system distance would have magnitude
$R\lapp 15$. The inspection of archive images (ESO-DSS)
 shows that no such bright stars are observed close 
 to the nominal position of \psr.
On the other hand the very small observed eccentricity of the orbit ($e = 5\times
10^{-5}$; L06) strongly argues against the hypothesis that the
companion is a NS. In fact, in this case it should be the remnant
of the massive star that first recycled \psr, and then exploded in a
supernova. This should have probably left a significant eccentricity in
the system, at odds with the observations.

In view of the considerations above, the companion is most likely a
massive WD, and the system J1439-5501 belongs to the growing class of the
so-called IMBPs (Camilo et al. 1996). 
Given its orbital and spin period, the favored
scenario, among those discussed by Tauris et al. (2012), 
is that the system was originated from the evolution of an IMXB with an asymptotic
giant branch companion, through a common envelope phase.

\section{Observations and data analysis}
\label{Sec:analysis}

The photometric data set used for this work consists of a series of
ground-based optical images acquired with the FOcal Reducer/low
dispersion Spectrograph 2 (FORS2) mounted at the ESO-VLT.  We
performed the observations in the {\it Standard Resolution mode}, with a
pixel scale of $0.126\arcsec /pixel$ (adopting a binning of $1\times1$ pixels) 
and a field of view (FOV) of $6'.8\times6'.8$.  All the brightest stars in
the FOV have been covered with occulting masks in order to avoid
artifacts produced by objects exceeding the detector saturation limit
in long exposures, which would have significantly hampered
the search for faint objects.

A total of 39 deep images in the $B_{HIGH}$, $V_{HIGH}$ and $I_{BESS}$
bands were collected during five nights
in May 2009, under program $383.D-0406(A)$ (PI: B. Lanzoni).  Since
the goal of this work is to identify the companion to \psr, only the
chip containing the region around the nominal position of the PSR has been analyzed.

By following standard reduction procedures, we corrected the raw
images for bias and flat-field.  In particular, in order to obtain
high-quality master-bias and master-flat images, 
we combined a large number of  BIAS and FLAT images obtained 
during the observation period by using the tasks {\tt zerocombine}
and {\tt flatcombine} in the  IRAF\footnote{ IRAF is distributed by the National Optical
 Astronomy Observatory,
which is operated by the Association of Universities for Research in Astronomy, Inc., under the cooperative 
agreement with the National Science Foundation.} package {\sc ccdred}. The
calibration files thus obtained have been applied to the raw images by
using the dedicated task {\tt ccdproc}.

We carried out the photometric analysis by using {\sc daophot}
(Stetson 1987, 1994).  The point spread function (PSF) has been
modeled in each image by using about 100 bright, isolated and not
saturated stars.  The PSF model and its parameters have been chosen
using the {\sc daophot} {\tt PSF} routine on the basis of a $\chi^2$
test. A Moffat function (Moffat 1969) turned out to provide the best fit
in all images.

Since our purpose is to detect the faintest stars in the field, we first
combined all the available images obtaining a master frame. Then we imposed a detection
limit of $3\sigma$, where $\sigma$ is the standard deviation of the measured background. 
By using the {\sc daophot} {\tt FIND} routine, we thus obtained a master list of
objects.
Finally, the star positions in this master frame has been adopted 
as reference to force the object detection and PSF fitting in each single image,
by using {\tt allframe} (Stetson 1987, 1994).
This procedure also allowed us to achieve
an improved determination of the star centroids and a better
reconstruction of the star intensity profiles.  At the end of the
reduction procedure we obtained a catalog of about  4000 sources (see Table \ref{Table:cat} for a small sub-sample of the catalog).

{\it Photometric calibration:} We selected ten bright and isolated
stars and, for each of them, we performed aperture photometry with
different radii ($r$) and we compared these magnitudes with those obtained
with the PSF fitting.  The mean value  of the differences between PSF and
aperture magnitudes has been found to be constant for $r\ge13$ pixels.  Thus we used
the value at $r=13$ pixels  as aperture correction to be
applied to all the stars in our catalog.  For a straightforward
comparison with theoretical models, we decided to
calibrate the instrumental magnitudes ($b$, $v$, $i$) to the standard Johnson
photometric system ($B$, $V$, $I$).  To this aim, we first derived the calibration
equations for ten standard stars in the field PG1323 (Stetson 2000),
which has been observed with FORS2  during  the observing run under photometric conditions.
To analyze the standard star field we used the {\sc daophot} {\tt PHOT}
task and we performed aperture photometry with the same radius used for
the aperture correction.  We then compared the obtained magnitudes
with the standard Stetson catalog available on the CADC web
site\footnote{http://cadcwww.dao.nrc.ca/community/STETSON/standards/}.
The comparison shows a clear dependence on color. Hence we
performed a linear fit in order to derive the trend as a function of the
color $(v-i)$ in case of $V$ and $I$ bands and of the $(b-v)$ for the
$B$-band.  The resulting calibration equations are
$B=b+0.126(b-v)+27.16$,
$V=v+0.051(v-i)+27.53$ and $I=i-0.002(v-i)+26.95$  and the final 
uncertainties on the calibrated magnitudes  are 
$\pm0.016$, $\pm0.004$ and $\pm0.010$ for $B$, $V$ and $I$, respectively.  We neglected the dependence 
on airmass since all exposures were taken at similar values.

{\it Astrometry:} Since in the Galactic field proper motions may
not be negligible, 
we used as astrometric reference stars the objects in the catalog
PPMXL (Roeser, 2010), where the proper motion of
each star is listed. Since the number of objects in common with our dataset is
large enough (126 stars), we could derive appropriate coordinate
transformations.  As first step of the procedure we derived the
position of the astrometric stars at the epoch of the observations.
Then we registered the pixel coordinates of the reference image onto
the absolute coordinate system through the cross-correlation of the
primary astrometric standards in common with our catalog, by using
CataXcorr\footnote{CataXcorr is a code aimed at cross-correlating
catalogs and finding astrometric solutions, developed by
P. Montegriffo at INAF - Osservatorio Astronomico di Bologna.  This
package has been successfully used in a large number of papers of our group in the
past 10 years.}.   The root mean square of the adopted transformations is  
$\sim0.3\arcsec$ in right ascension ($\alpha$) and $\sim0.2\arcsec$ in declination ($\delta$), 
while the typical uncertainty of the PPMXL stars in this field is $\sim0.1\arcsec$. These quantities give an accuracy 
of $\sim0.11\arcsec$ for our astrometric solution.

\section{The identification of the companion to PSR J1439-5501}
\label{Sec:Results}

In order to identify the companion to \psr\ we focused our attention
to any object located close to the PSR nominal position, as derived
from the timing solution in the radio band: $\alpha_{2000}=14^{\rm h} 39^{\rm m}
39^{\rm s}.742(1)$ and $\delta_{2000}=-55^\circ 01' 23''.62(2)$ at the
reference epoch MJD=53200 (L06). No determination of the PSR proper
motion was reported in L06. However, assuming a conservative transverse velocity of
$\sim 100$ km $s^{-1}$ (Hobbs et al., 2005) the expected total positional shift would be $\sim
0.15\arcsec$ (for the system distance) over the $\sim 5$ yr
interval between the reference epoch of the radio ephemeris and the
date of optical observations. Therefore, the offset
between the position of the system J1439-5001
in the optical images and the coordinates given by L06
would be at most of the order of the astrometric solution accuracy.  
We also note that follow-up timing observations
of the  system provide additional
support to a safe use of the coordinates reported in L06: in fact
these observations indicate a transverse velocity for \psr\ well below
$100$ km s$^{-1}$ (Burgay, private communication).

\begin{figure*}
\includegraphics[width=150mm]{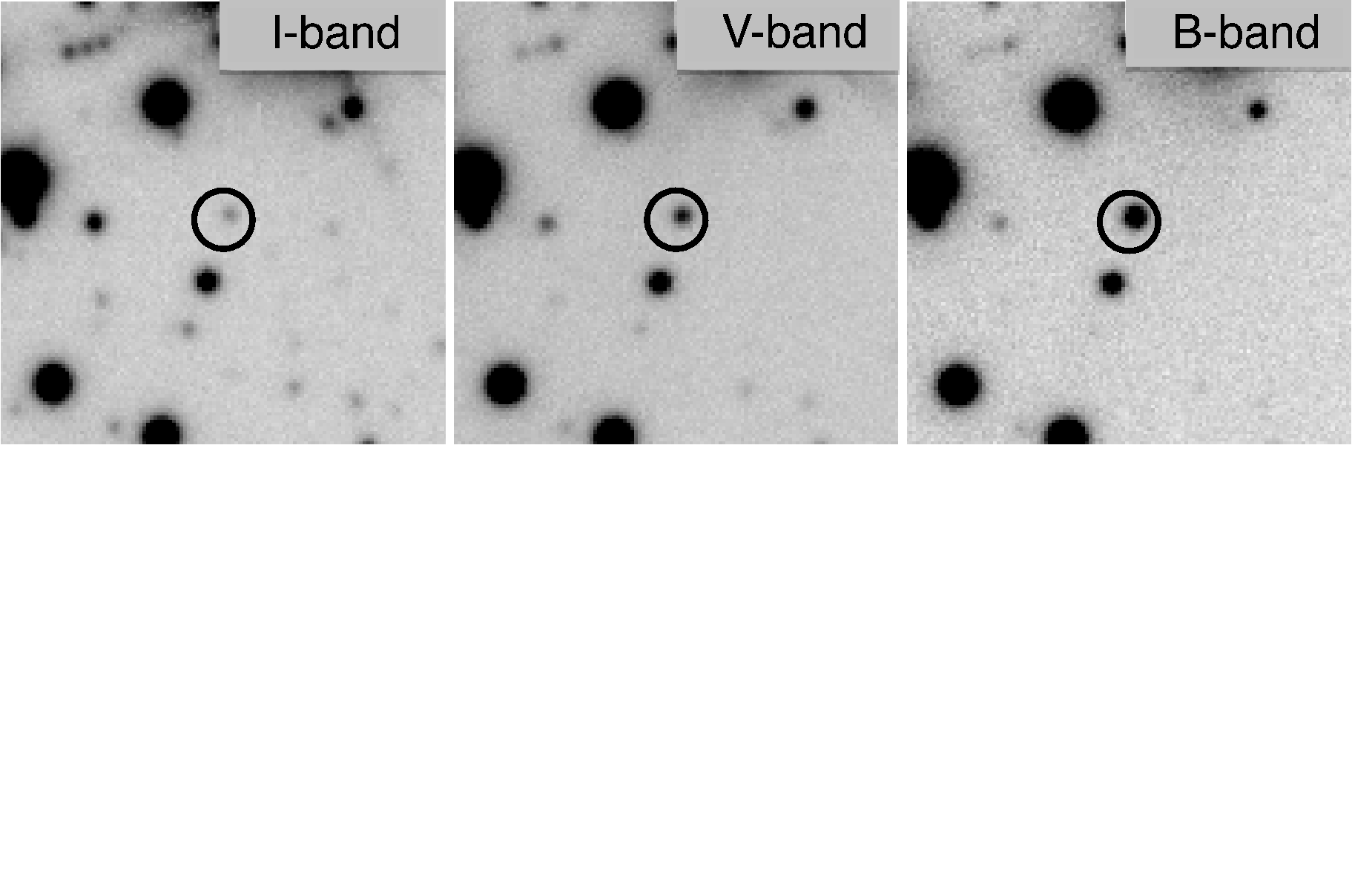}
  \caption{From left to right, $15\arcsec \times 15\arcsec$ maps in the $I_{BESS}$,  $V_{HIGH}$ and $B_{HIGH}$
  bands around the PSR nominal position. The black circles are centered on \psr\ 
 and have a radius of $1\arcsec$.}
\label{map}
\end{figure*}

A visual inspection of the deep images shows that there is a star located at
only $0.05\arcsec$  from the nominal
position of \psr\ (see Figure \ref{map}).  Moreover, no other star in our
catalog is found within an error circle 
centered on the PSR position and having a radius of  several
times the aforementioned uncertainty in the astrometric solution.

\begin{figure*}
\includegraphics[width=150mm]{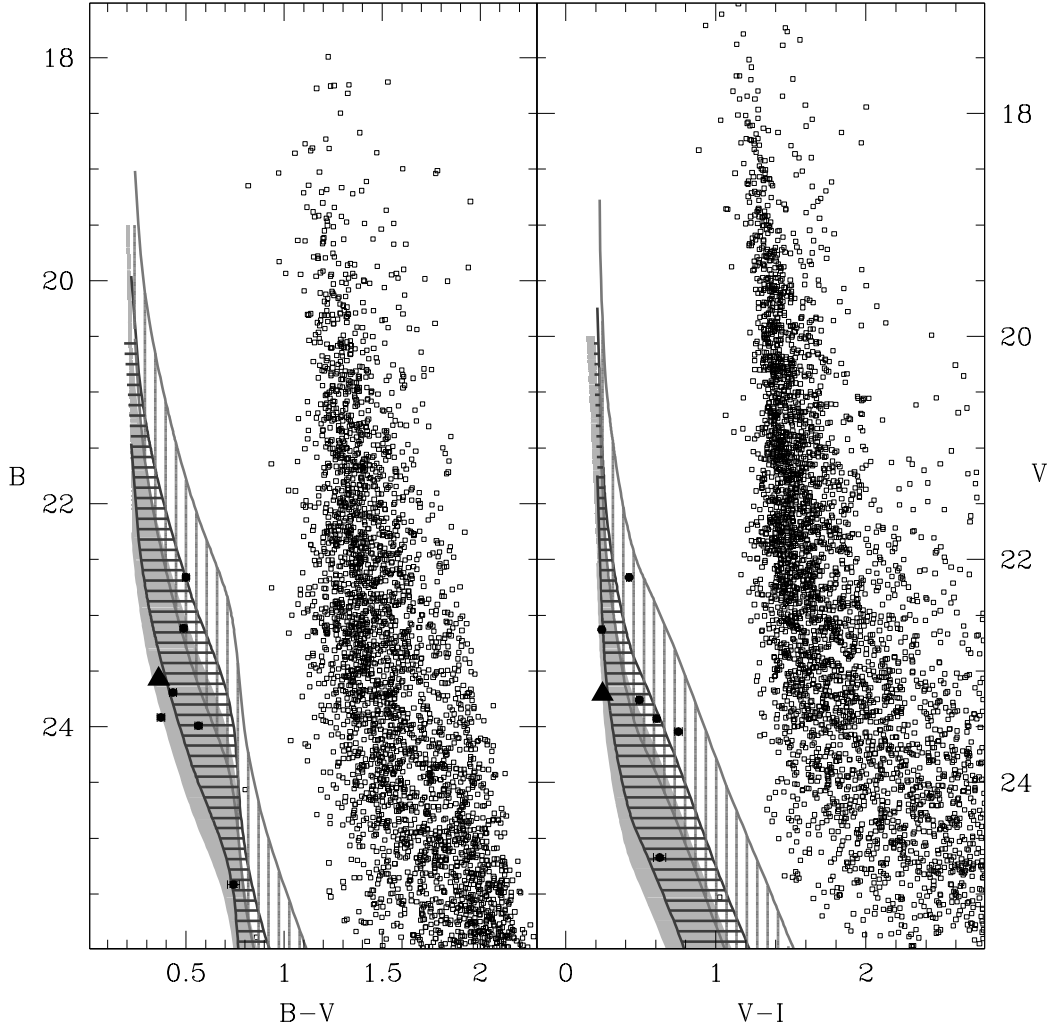}
  \caption{
  ($B$, $B-V$) and ($V$, $V-I$) Color-Magnitude Diagrams for
  \com\ to \psr\  (black triangle) and for all the other objects observed in the detector FOV.  The companion is  located at $B=23.57\pm0.02$,   $V=23.21\pm0.01$ and   $I=22.96\pm 0.02$, corresponding to the region of WD cooling sequences. The
   vertically and the horizontally hatched bands correspond to the CO-WD cooling sequences, respectively, for  $0.6$ and
  $1.0~\Msun$, in a range of
  distances between 600 and 1200 pc (BaSTI database; Salaris et al., 2010), while the shaded light gray strip marks, in the same distance range, the location of a $1.0~\Msun$ ONe-WD (Althaus et al., 2007).
}
\label{cmd}
\end{figure*}

A comparison among the $I$, $V$ and $B$ images  shown in Figure \ref{map}
clearly suggests that this star has a color bluer than most of others
objects in the field.  This feature is confirmed by the inspection of the
Color-Magnitude Diagrams (CMDs), where the star
 is located on the left side of the bulk of the detected stars, in a region compatible
with  WD cooling sequences (see Figure \ref{cmd}). This is in nice
agreement with the scenario proposed for the evolution of the system J1439-5501 
discussed in Section~\ref{Sec:pulsar}.
The probability that any star in our catalog
falls at the PSR position by chance coincidence is low ($\sim 2\%$), 
and it further reduces to $\sim 0.003\%$ if only WDs are
considered. The combination of all these  pieces of evidence strongly suggests
that the detected star is the companion to \psr.

 In principle,  companions to MSPs can show optical variability, due to irradiation by the PSR or by rotational modulation, as observed in the case of PSR 0655+64
(Van Kerkwijk et al., 1995 \& Van Kerkwijk, 1997).
By comparing the flux of the PSR intercepted by the WD with the WD flux we estimated that the flux enhancement due to re-heating is negligible.
Unfortunately we could not check if the system shows any variability due to rotation since 
the largest available number of observations is in the $I$ band, in which the object is very faint (see Figure \ref{map}), while on the $B$ and $V$ bands the orbital sampling is very poor and prevents any study of variability. 
Hence with the available data it is not possible to set a definitive 
conclusion about the presence of magnitude modulations. Therefore, in order to detect possible variability, and eventually study its nature, 
future phase resolved observations are required.
 
\begin{figure*}
\includegraphics[width=150mm]{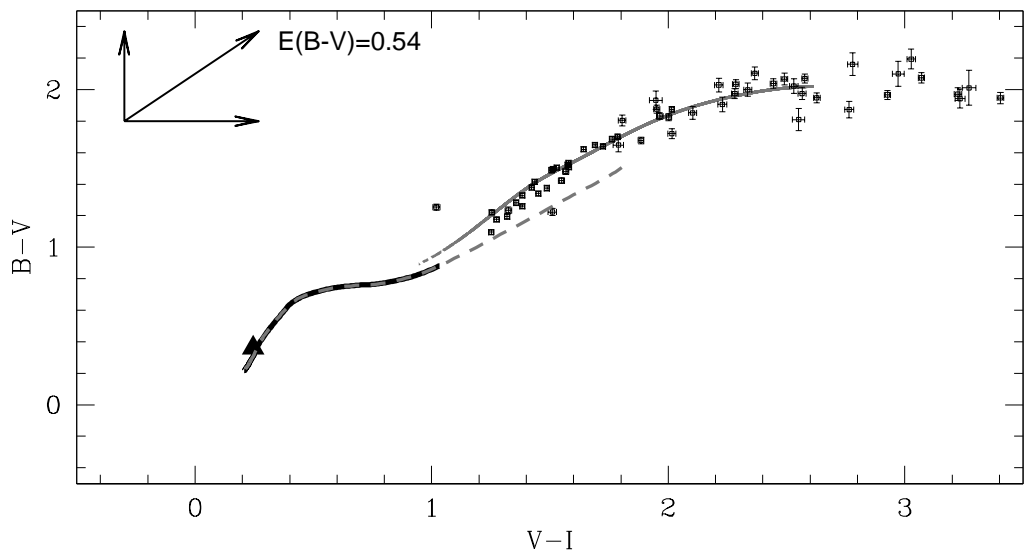}
  \caption{ ($V-I$, $B-V$) Color-Color diagram for all stars (open squares)  within $30\arcsec$ from \psr. 
   The companion
  is highlighted with a large black triangle.
  The gray  solid  region marks the location of MS populations having solar metallicity 
  and different ages, consistent with those observed in the   Galactic field. 
   The  black solid and gray dashed  lines  correspond to the cooling sequences for  $1.2 \Msun$  CO- and ONe-WDs, respectively. 
  A color excess $E(B-V)=0.54$ is applied to the models (arrows mark the entity and direction of the applied reddening).
  }
\label{colcolred}
\end{figure*}

With the aim of deriving mass and age of \com, we compared its position
 in the CMDs ($B=23.57 \pm 0.02 $, $V=23.21 \pm 0.01$ and $I=22.96 \pm 0.01$) with  a set  of theoretical CO- and ONe-WD cooling
sequences of different masses (BaSTI database, Salaris et al., 2010; Althaus et al., 2007).  
In order to make consistent the ONe-WD models with CO-WD cooling sequences we applied to the formers 
the same color transformations used for the latters.
 As a first step, we calculated the
extinction coefficient $E(B-V)$ by generating the  Color-Color (CC) diagram
 ($V-I,B-V$),  where  the dependence on distance disappears.
 In particular, we compared the observed distribution of MS stars (within $30\arcsec$ from \psr) with the locus of theoretical models, with 
solar metallicity and different ages, typical of the Galactic field population.
Different reddening values,  ranging  between 0.4 and 0.7 with steps of $\delta E(B-V)=0.001$, have been iteratively applied to theoretical models.
By performing a  $\chi^2$ test, we obtained that the best-fit value\footnote{The uncertainties on $E(B-V)$ have been estimated by accounting for the   magnitude errors of the observed population.} is $E(B-V)\sim0.54^{+0.06}_{-0.05}$ (see Figure~\ref{colcolred}).

In the following we will adopt $E(B-V)=0.54$, keeping in  mind that any estimate 
of the WD mass, distance and derived quantities depends on the $E(B-V)$ value.
As can be seen from  Figure~\ref{colcolred},  if   $E(B-V)=0.54$   is assumed, 
the colors of \com\ nicely match the theoretical sequences  for WDs\footnote{ Note 
 that the  theoretical models shown in Figure~\ref{colcolred} are for  $1.2~\Msun$ WDs.
 However  the dependence of the WD cooling sequences on mass is negligible 
in the ($V-I,B-V$) CC Diagram (Bergeron et al., 1995).}. 

 By taking into account  the absorption coefficients properly calculated for the effective wavelengths 
of the filters  (Cardelli et al., 1989),
we derived the unabsorbed colors $(V-I)_0=-0.32^{+0.09}_{-0.07}$ and $(B-V)_0=-0.21^{+0.09}_{-0.07}$,   
 from which we estimated a temperature  $T=31350^{+21500}_{-7400}$.

 By using the derived value of the reddening and  the adopted range of distances ($600~-~1200$ pc; see section {\ref{Sec:pulsar}}) 
we placed the WD cooling sequences for masses of $0.6$ and $1.0~\Msun$ selected from the BaSTI database (Salaris et al. 2010)  in the ($V, V-I$) and the ($B, B-V$) CMDs.
 As can be seen from Figure~\ref{cmd}, 
\com\ is clearly not compatible with low mass WDs, while it 
appears to be slightly more massive than $1.0~\Msun$. 

\begin{figure*}
\includegraphics[width=150mm]{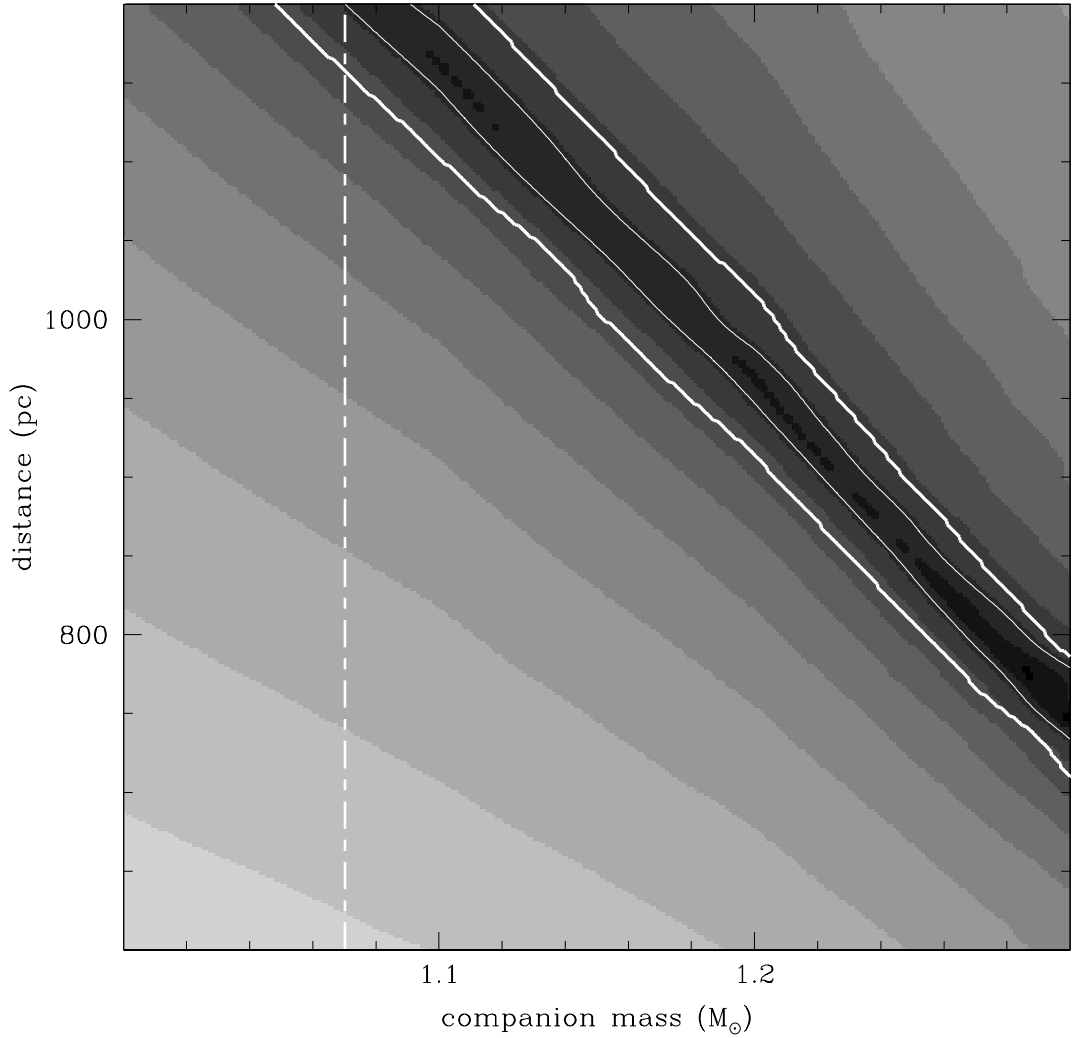}
  \caption{Distribution of probabilities  for CO-WDs in the plane of companion mass ($m$) and system distance ($d$). Darker regions mean more probable configurations. 
  The thick white contour marks the region occupied by couples $m$-$d$ providing confidence values $\le1$ in both CMDs, 
  the thin white line encircles the configurations with confidence value $\le 1$.}
\label{mapmd}
\end{figure*}

In order to better constrain the mass of \com\ we needed a tight sampling in mass.
Starting from the available tracks for CO- and ONe-WDs in the mass range $1-1.25~\Msun$ (Salaris et al., 2010; Althaus 2007)
and assuming as a first approximation a linear relation between magnitude, color and mass in such a small range, 
we derived cooling sequences  between $1.0$ and $1.25~\Msun$ at regular steps.
By using the same linear relation we obtained tracks up to $1.3~\Msun$. 
For each mass ($m$) step we varied the PSR distance ($d$)  in the  range $600-1200$ pc,
and we calculated the difference ($\Delta$ expressed in magnitude units) between 
the observed location of \com\ in the CMD and its perpendicular projection onto the cooling sequence.
We applied this method in both the ($B,B-V$) and ($V,V-I$) CMDs,
thus obtaining $\Delta_1$ and $\Delta_2$, respectively, for any $m$-$d$ pair.
For each value of $m$ we then selected the value of $d$ which minimizes the sum of  $|\Delta_1|+|\Delta_2|$
and we associated a confidence value to each $m$-$d$ pair.
In each CMD the confidence value is 
calculated as $|\Delta|$ normalized to the photometric combined error $\sqrt{(e_{COL}^2+e_{MAG}^2)}$ 
of the star. In this way, the smaller is the confidence value, the larger is the probability 
associated to that configuration. In particular, a confidence value $\le 1$ means that 
the cooling sequence and the observed position are in agreement within the
photometric errors. The resulting confidence value, when information from both CMDs is combined, 
is the sum of the confidence values derived from the two CMDs.
The resulting distribution of confidence values in the $m$-$d$ plane is plotted in Figure~\ref{mapmd}. 
All the $m$-$d$  couples for which the confidence value is $\le1$ in both CMDs, have
been selected as plausible combinations of $m$ and $d$ for \com\ 
(they are encircled by the white contour in Figure~\ref{mapmd}).

From this analysis we find that, for the considered upper limit to the distance 
($d=1200$ pc), $M_{\rm COM} \gapp 1 \Msun$ (see Figure~\ref{mapmd}). 
Considering the constraint on the minimum companion mass from the PSR mass function ($M_{\rm COM}>1.07 \Msun$), we can safely confirm that
$d\le1200$ pc.
At small distances, the analysis is limited by the availability of WD models (see above).
However, the adopted upper limit for the mass in our analysis ($M_{\rm COM}=1.3\Msun$) likely does not affect the results, even if we cannot definitely  rule out a more massive companion in the range $1.3<M_{\rm COM}<M_{\rm CH}$, where $M_{\rm CH}=1.44\Msun$ is the Chandrasekhar mass limit for a WD (Chandrasekhar, 1935).
If \com\ is  a CO-WD  it would have a mass $1.07 \lapp M_{\rm COM}  \lapp 1.3~\Msun$, located at a distance $710\lapp d \lapp 1200$ pc (see Figure~\ref{mapmd}), which imply a radius in the range $4.5-7.7~10^{-3}~\Rsun$.  
In case of ONe-WDs models, we found $m$-$d$  configurations as probable as those obtained for CO-WDs. Hence, \com\ to \psr\ could also be a ONe-WD with mass $1.07\lapp M_{\rm COM} \lapp 1.3~\Msun$, radius in the range $4.0-7.8~10^{-3}~\Rsun$ and located at distance  $640\lapp d \lapp1200$ pc. 
Note that, because of the larger molecular weight of the ONe-WD with respect to the CO-WDs, the former have a smaller radius and hence, for a fixed mass, to fit the observed properties they should be located closer than CO-WDs.

\section{Discussion and conclusions}
\begin{figure*}
\includegraphics[width=150mm]{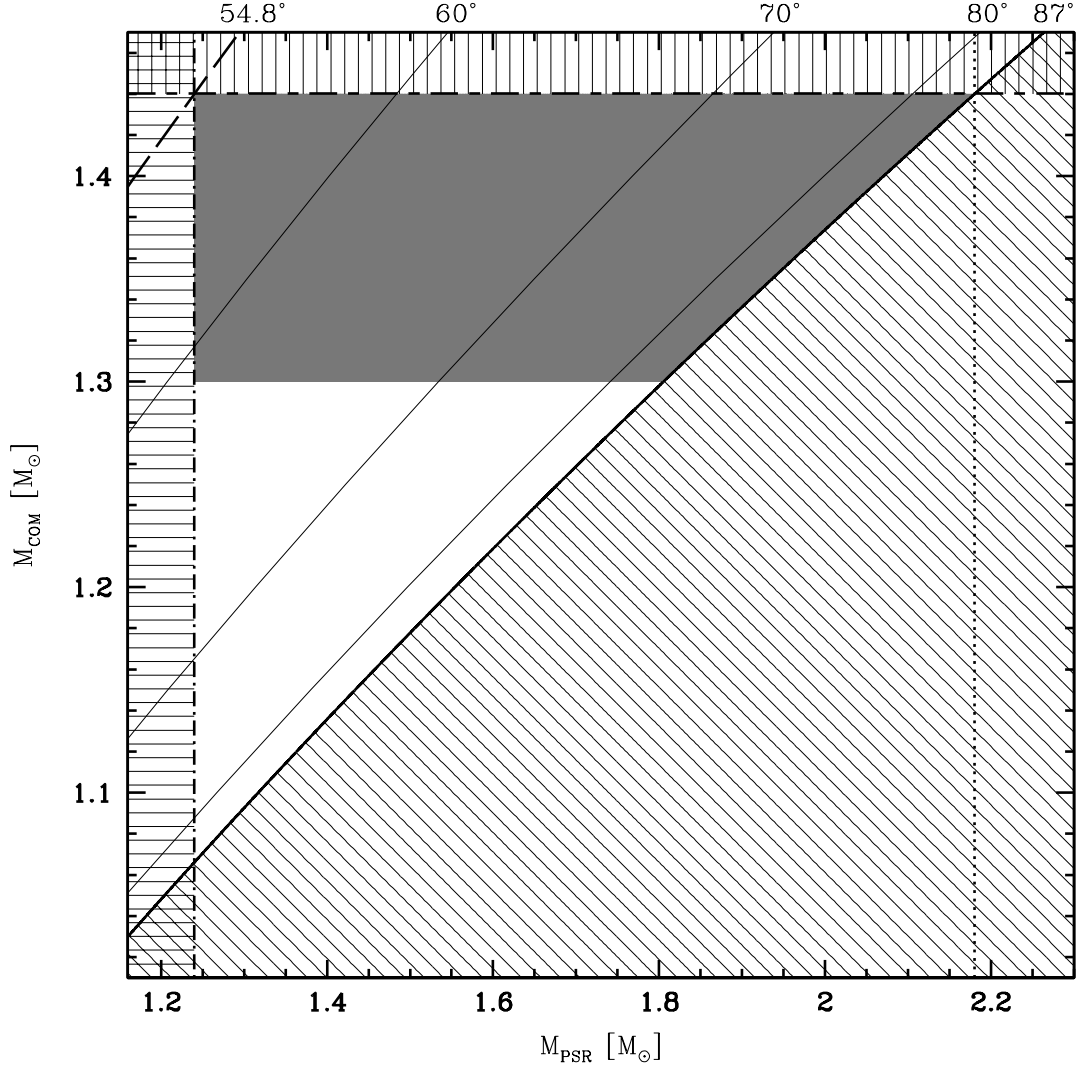}
  \caption{
  Companion mass as a function of the PSR mass.
Solid lines indicate constant orbital inclinations from the PSR mass function.
In particular the thick solid line marks the limit  of $i_{\rm max}=87^{\circ}$.
The short-long dashed line marks the physical limit $M_{\rm CH}=1.44 \Msun$ for the WD mass.
The dot dashed line marks the minimum limit $M_{\rm NS,min}=1.24\Msun$ ever measured  for a radio pulsar mass. 
 The hatched regions are excluded because of the combination of the physical limits above.
The dotted line indicates the upper limit to the PSR  mass $M_{\rm PSR}<2.18$ derived assuming $M_{\rm COM}=M_{\rm CH}$ and $i=i_{\rm max}$; while the long dashed line indicates the lower limit to the inclination angle $i>54.8^{\circ}$ calculated assuming $M_{\rm COM}=M_{\rm CH}$ and $M_{\rm PSR}=M_{\rm NS,min}$.
The white region marks the investigated range of WD masses by comparison between optical photometry and theoretical models. 
Note that we could not extend the analysis to  companions more massive than $1.3 \Msun$ (gray region) because models for WD with larger  masses are not available.}
 \label{mapmpmc}
\end{figure*}

From the PSR mass function  $f_{\rm PSR}=0.227597~\Msun$ (L06) and the 
range of permitted companion masses and radii derived in the previous Section, 
we constrained $\mpsr$ and the inclination of the system, as summarized in Figure \ref{mapmpmc}.
First of all, given the relatively large orbital separation $\sim 9.5~\Rsun$
and the small radius of the companion, the absence of observed
eclipses of the radio signal along the orbit only constraints
$i<89.4^\circ$.  
The observation of the Shapiro delay effect could allow one to impose further constraints on the inclination. Unfortunately, the quality of the available radio timing (with a non uniform orbital coverage especially close to superior conjunction, where the effect of the Shapiro delay is maximum) does not allow at the moment to extract a constraining determination of the Shapiro delay parameters in this system. However, simulations show that, even with the few times of arrival that we have around orbital phase 0.25, a signature of an almost edge-on orbit would be detectable. We can hence constrain the inclination angle to be $\lapp 87^\circ$.
Simulating a data set with monthly observations and the current instrumentation, a determination of the Shapiro delay would take from $\sim 1$ to 3 decades, depending on the inclination of the source (in the range from 60 to 85 degrees).
Instead, a lower limit of $54.8^\circ$ for $i$ results from the assumption that $\mcom=M_{\rm CH}$ and that the PSR has a mass larger than $1.24 \Msun$ (see Section \ref{Sec:intro}).
Moreover, by adopting $\mcom=M_{\rm CH}$ and $i=87^{\circ}$, an upper 
limit of $2.18~\Msun$  can be inferred for the PSR mass.  

Of all the CO-WD companions to IMBPs identified so far,  \com\ to \psr\  is one of the few with a 
mass estimate.
Moreover, comparing its estimated mass with the median  of companion masses of previously identified CO-WDs
(see Tauris et al. 2012),  it turns out to be among the most massive.

Although the cooling age of a WD suffers from uncertainties, it is the only reliable age indicator 
of a PSR binary system (Tauris, 2012).
Hence, the optical identification of a WD companion to a binary millisecond pulsar and its cooling age estimate 
are of crucial importance to constrain the spin-down theory 
and understand how the characteristics and the age of a PSR are related.
Kulkarni (1986) proposed for the first time a comparison between these two ages,
finding a quite good agreement.
However, there is an
increasing body of evidence (see e.g. Tauris et al., 2012) that the
spin-down age is a very poor  measure of the time that a recycled PSR
spent since the completion of mass transfer. 
In fact,  in several PSRs, WDs are observed to be
both older and younger than the millisecond pulsar (Hansen \& Phinney, 1998a,b).
However, in most cases the NS seems to be older than the WD because the standard spin-down model may 
overestimate the PSR age (Jacoby et al., 2006). 
For the parameters inferred above, the cooling age 
of \com\ is in the range   $ 0.1\lapp t_{cool} \lapp 0.2$ Gyr and $0.1\lapp t_{cool} \lapp 0.4$ Gyr for CO- and ONe-WD respectively ( and up to $0.4-0.5 $ Gyr for the minimum reddening configuration) and it is several times smaller than its estimated spin-down age ($3.2-4.5$ Gyr; Kiziltan and Thorsett, 2010).
This discrepancy is not surprising. In fact the characteristic age for PSRs in the millisecond regime could not be a proper estimate of the PSR ages,
since the hypothesis that $P_0$ is negligible with respect to the current $P$
is not applicable to recycled PSRs.
Such a scenario is in agreement with the results of simulations of  an evolved synthetic population of MSPs (Kiziltan and Thorsett, 2010), 
which shows that the characteristic ages could either over- or under-estimate the true age of MSPs by more than a factor of ten.
Hence, we stress how  the identification of the companion and its age estimate (e.g. from the cooling sequences for WD companions) could be a powerful tool to derive the true age of a recycled PSR.

 While a spectroscopic analysis, both for  the radial velocity curve and the chemical analysis, is hardly feasible with the current generation of instrumentation, a photometric follow-up could provide plenty of useful information. In particular, phase resolved datasets could lead to the possibility of revealing any optical variability of the companion, while multi-band photometry could allow to better estimate the reddening and, by applying the same method used in this paper, to better constrain the mass and the distance of the companion.

\section{Acknowledgement}
We thank the Referee Van Kerkwijk for the careful reading of the manuscript and the useful comments.
This research is part of the project Cosmic-Lab funded by the European Research Council 
(under contract ERC-2010-AdG-267675).

\newpage

\newpage

\newpage

\newpage

\newpage

\newpage
\begin{table}
\begin{center}
\begin{tabular}{| l | c  c | c  c | c  c | c  c | }
\hline
\hline
ID &  R.A. (J2000) & Dec. (J2000) & $B$ & $eB$ & $V$ & $eV$ & $I$ & $eI$\\
\hline
     1$\star$ &   14:39:39.746   &  -55:01:23.66   &     23.57   &  0.02    &  23.21  &   0.01  &  22.96   &   0.02\\
     2   &   14:39:43.006   &  -55:00:47.58   &     22.24   &  0.02    &  21.00  &   0.01  &  19.66   &   0.01\\
     3   &   14:39:41.277   &  -55:01:10.81   &     22.16   &  0.02    &  20.51  &   0.01  &  18.82   &   0.01\\
     4   &   14:39:40.619   &  -55:01:14.66   &     21.46   &  0.02    &  19.96  &   0.01  &  18.43   &   0.01\\
     5   &   14:39:40.234   &  -55:00:55.79   &     22.70   &  0.02    &  21.21  &   0.01  &  19.71   &   0.01\\
     6   &   14:39:38.596   &  -55:00:40.84   &     21.34   &  0.02    &  19.98  &   0.01  &  18.60   &   0.01\\
     7   &   14:39:38.068   &  -55:00:45.54   &     23.09   &  0.02    &  21.75  &   0.01  &  20.32   &   0.01\\
     8   &   14:39:37.287   &  -55:00:50.09   &     23.53   &  0.02    &  21.98  &   0.01  &  20.28   &   0.01\\
     9   &   14:39:36.626   &  -55:01:24.52   &     21.39   &  0.02    &  20.20  &   0.01  &  18.88   &   0.01\\
    10  &   14:39:35.910   &  -55:01:29.44   &     23.68   &  0.02    &  22.14  &   0.01  &  20.49   &   0.01\\
\hline
\end{tabular}
\end{center}
\caption{Position and $B$, $V$ and $I$ magnitudes (with relative errors) of ten stars around \com\ to \psr. The $\star$ marks the identified companion. }
\label{Table:cat}
\end{table}

\end{document}